\documentclass[conference]{IEEEtran}
\IEEEoverridecommandlockouts
\usepackage{cite}
\usepackage{amsmath,amssymb,amsfonts}
\usepackage{algorithmic}
\usepackage{graphicx}
\usepackage{textcomp}
\usepackage{xcolor}
\def\BibTeX{{\rm B\kern-.05em{\sc i\kern-.025em b}\kern-.08em
		T\kern-.1667em\lower.7ex\hbox{E}\kern-.125emX}}
\begin{document}
\title{Arrhythmia Classifier Using Convolutional Neural Network with Adaptive Loss-aware Multi-bit Networks Quantization}

\author{
	\IEEEauthorblockN{1\textsuperscript{st} Hanshi Sun}
	\IEEEauthorblockA{\textit{School of Electronic Science and}\\\textit{Engineering} \\
		\textit{Southeast University}\\
		Nanjing, Jiangsu, China \\
		preminstrel@gmail.com}
	\and
	\IEEEauthorblockN{2\textsuperscript{nd} Ao Wang}
	\IEEEauthorblockA{\textit{School of Electronic Science and}\\\textit{ Engineering} \\
		\textit{Southeast University}\\
		Nanjing, Jiangsu, China\\ 
		Ao0323@outlook.com}
	\and
	\IEEEauthorblockN{3\textsuperscript{rd} Ninghao Pu}
	\IEEEauthorblockA{\textit{School of Electronic Science and}\\\textit{ Engineering} \\
		\textit{Southeast University}\\
		Nanjing, Jiangsu, China \\
		1742237531@qq.com}
	\and
	\IEEEauthorblockN{4\textsuperscript{th} Zhiqing Li}
	\IEEEauthorblockA{\textit{School of Electronic Science and}\\\textit{ Engineering} \\
		\textit{Southeast University}\\
		Nanjing, Jiangsu, China \\
		lizhiqing@seu.edu.cn}
	\and
	\IEEEauthorblockN{5\textsuperscript{th} Junguang Huang}
	\IEEEauthorblockA{\textit{School of Electronic Science and}\\\textit{ Engineering} \\
		\textit{Southeast University}\\
		Nanjing, Jiangsu, China \\
		huangjunguang@seu.edu.cn}
	\and
	\IEEEauthorblockN{6\textsuperscript{th} Hao Liu\textsuperscript{\dag}}
	\IEEEauthorblockA{\textit{National ASIC System Engineering}\\\textit{Research Center} \\
		\textit{Southeast University}\\
		Nanjing, Jiangsu, China \\	
		nicky\_lh@seu.edu.cn}
	\and
	\IEEEauthorblockN{7\textsuperscript{th} Zhi Qi\textsuperscript{\dag}}
	\IEEEauthorblockA{\textit{National ASIC System Engineering}\\\textit{Research Center} \\
		\textit{Southeast University}\\
		Nanjing, Jiangsu, China \\	
		101011256@seu.edu.cn}
}
%
%
%
%
%
\maketitle

\begin{abstract}
	Cardiovascular disease (CVDs) is one of the universal deadly diseases, and the detection of it in the early stage is a challenging task to tackle. Recently, deep learning and convolutional neural networks have been employed widely for the classification of objects. Moreover, it is promising that lots of networks can be deployed on wearable devices. An increasing number of methods can be used to realize ECG signal classification for the sake of arrhythmia detection. However, the existing neural networks proposed for arrhythmia detection are not hardware-friendly enough due to a remarkable quantity of parameters resulting in memory and power consumption. 
	In this paper, we present a 1-D adaptive loss-aware quantization, achieving a high compression rate that reduces memory consumption by 23.36 times. In order to adapt to our compression method, we need a smaller and simpler network. We propose a 17 layer end-to-end neural network classifier to classify 17 different rhythm classes trained on the MIT-BIH dataset, realizing a classification accuracy of 93.5\%, which is higher than most existing methods. Due to the adaptive bitwidth method making important layers get more attention and offered a chance to prune useless parameters, the proposed quantization method avoids accuracy degradation. It even improves the accuracy rate, which is 95.84\%, 2.34\% higher than before. Our study achieves a 1-D convolutional neural network with high performance and low resources consumption, which is hardware-friendly and illustrates the possibility of deployment on wearable devices to realize a real-time arrhythmia diagnosis.
\end{abstract}
\begin{IEEEkeywords}
	arrhythmia, neural network, quantization, multi-bit networks, multi-class classification, machine learning, resource-constrained devices
\end{IEEEkeywords}

\section{Introduction}
In many healthcare scenarios, patients are diagnosed with a remarkable variety of diseases, including Cardiovascular disease (CVDs), a universal deadly disease \cite{Serhani2020}. The electrocardiogram (ECG) depicts the human heart's electrical activity and is significant for accurate diagnoses. However, in the early stage, with unobvious symptoms and short duration, some arrhythmias may be challenging to recognize \cite{Krishnakumar2020},  resulting in severe consequences. Therefore, real-time heart rate detection deployed on low-power devices have come under the spotlight.

Neural networks establish a mapping from low-level signals to high-level semantics by simulating a hierarchical structure similar to the human brain to achieve the hierarchical feature expression of data, which has powerful information processing capabilities,  promoting the development of algorithms and models for ECG classification methods \cite{Liu2019}. Although the detection and classification accuracy of the neural network model seems considerable \cite{yildirim2018arrhythmia}, its huge trainable network parameters consume a large amount of memory and require more time for complex computation, which makes it difficult to deploy on low-power hardware platforms.

To tackle this issue, we consider both the design of network structure and the adaptation of quantitative compression method, which can reduce the accuracy degradation from typical quantization methods, even improve the accuracy in that model error is optimized by cited adaptive bitwidth quantization method. The contribution of this paper has three aspects:
 \begin{itemize}
 	\item An adaption of cited adaptive loss-aware quantization(ALQ) is proposed to lower the memory and power consumption of a 1-D convolutional neural network while maintaining or even improving the classification accuracy.
	 \item Based on our novel compression method, a 17 layer convolutional neural network (CNN) architecture for cardiac arrhythmia (17 classes) detection based on long-duration ECG fragment analysis is proposed, and it realizes an overall accuracy of 93.5\% for arrhythmia detection.
	 \item Finally, we implement the quantization method and achieve a classification accuracy of 95.84\% with a memory compression of 23.4 times, illustrating the superiority of the proposed quantization method over previous methods.
\end{itemize}

\section{Related Work}
Pattern recognition is widely used to automatic arrhythmia diagnosis \cite{park2008hierarchical, llamedo2010heartbeat}. By manually transforming the input into features that carry valuable cardiological knowledge, they can achieve notably high accuracy using a classifier for diagnosis. However, such performance always means a cost of human resources to realize hand-crafted feature extraction and have poor generalizability, relying on the features heavily.

Neural networks fuse the feature extraction and classification so that the arrhythmia class can be directly drawn by inference using raw ECG signal, achieving an end-to-end detection.
Y{\i}ld{\i}r{\i}m et al. propose a new wavelet sequence model based on a deep bidirectional LSTM network, classifying regular heart rate and six other types in the MIT-BIH arrhythmia database. The results show that the recognition performance of the modified model is up to 99.39\% \cite{yildirim2018novel}. Kiranyaz et al. employ a 3-layer CNN and 2-layer multi-layer perceptron to learn features of heartbeats \cite{Kiranyaz2015}.
The convolutional neural network was used to automatically detect normal and MI ECG beats (noisy and silent), with the average accuracy of noisy and denoised ECG beats reaching 93.53\% and 95.22\%, respectively \cite{acharya2017application}.
The Pan-Tompkins algorithm was employed to segment heartbeats and implemented a 5-layer DNN for arrhythmia detection \cite{pan1985real}. The former networks have its limitation that the input is required to be a fixed length, which compels the ECG signals to need to be divided into the same length as input. 

Y{\i}ld{\i}r{\i}m et al. used a 1-D CNN for a 10 s ECG signal, realizing a genuinely end-to-end diagnosis \cite{yildirim2018arrhythmia}, whose performance is better.  Hannun et al. develop a DNN classification of 12 heart rate levels using a patient's single-lead electrocardiogram, whose sensitivity exceeds the expert average. Therefore, deep neural networks (DNN) can classify different arrhythmias in single-lead ECG \cite{hannun2019cardiologist}. Although the accuracy is high, the neural networks are computationally intensive and consume remarkable memory, making it challenging to deploy deep neural networks on resource-constrained devices. For instance, the memory of an existing ECG classifier \cite{yildirim2018arrhythmia} is about 7852 KB, having difficulty being deployed on resource-constrained devices. Let alone some giant networks like ResNet-152, which has a size of 230 MB. It is almost impossible to realize real-time detection with such size. Therefore,  we should reduce their complexity to achieve the deployment on mobile and embedded devices. Without a doubt, substantial efforts should be made for speed-up and compression.

In order to take advantage of the pre-trained ECG signal classifier for efficient inference on resource-constrained devices, compression can be realized via pruning \cite{han2015deep}, quantization \cite{gong2014compressing, guo2017network}, distillation \cite{hinton2015distilling}. Compared with pruning and distillation, quantization compression should have a better performance due to the particularity of ECG signal structure coding. Vanhoucke et al. propose that neural networks have little accuracy degradation with 8-bit weights \cite{vanhoucke2011improving}. Qu et al. focus on quantizing the full precision weights of a DNN into binary encodes and the corresponding scaling factors \cite{courbariaux2015binaryconnect,qu2020adaptive}. Although these methods were originally aimed at image classification problems, they can be combined with CNN for arrhythmia diagnosis as well, which have a similar structure and mechanism.

\section{Methodology}
In this section, we introduce the architecture overview of the classifier firstly, and we describe the details of our 1-D CNN architecture. At the end of this section, the ALQ strategy and the choice of some ALQ parameters are discussed.

The overall proposed framework can be divided into two parts, as seen in Fig. 1. The first part is the arrhythmia classification neural network architecture, which is based on the basic block design and determines the depth of the neural network. After training the model, we can get a full precision ECGNet that achieves an accuracy of 93.5\%. The model parameters should be saved for the compression in the next part. 
\begin{figure}[h]
	\centering
	\includegraphics[width=\linewidth]{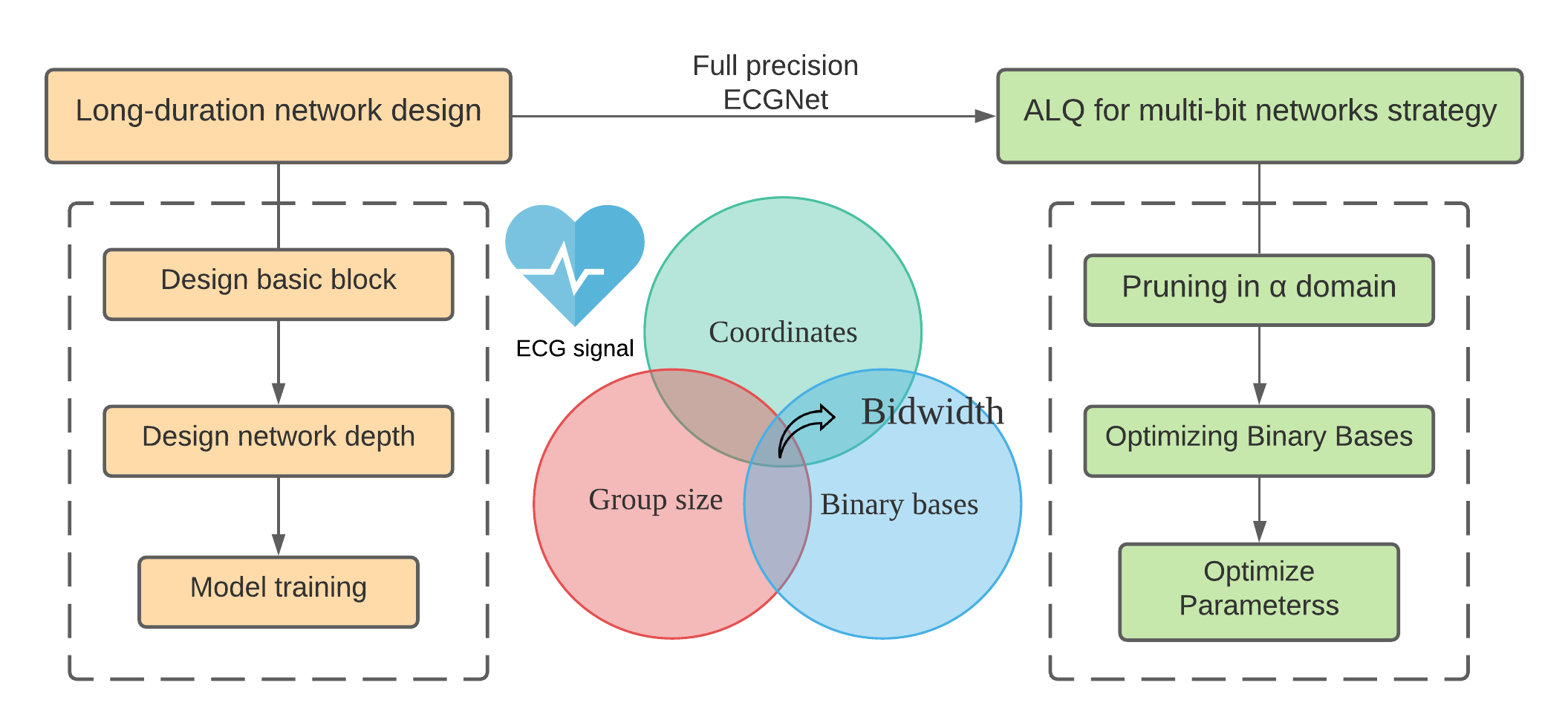}
	\caption{Overview of the proposed framework}
	\label{fig:Flowchart of the proposed framework}
\end{figure}

The second part is ALQ strategy (Adaptive Loss-aware Quantization). The sensitivity of each layer in the network to the quantization is different. Therefore, assuming that the total number of bits we give remains unchanged, the layers that are more sensitive to quantization have more bits, and the less sensitive layers have fewer bits, so as to achieve better accuracy, which reduces the average bitwidth by pruning the least significant coordinates in $\boldsymbol{\alpha}$ domain and optimize the binary bases $\boldsymbol{B}_k$ and coordinates $\boldsymbol{\alpha}_k$, basing on the correct choice of parameters such as $n$. This part realizes the powerful compression of the neural network, unlike existing methods, successfully avoids accuracy degradation, which can meet a lower resource requirement.

\subsection{Design of the 17 Layer CNN Using Long-duration ECG Fragments}

Our original arrhythmia classification convolutional neural network is presented in Fig. 2. The network is composed of a number of basic blocks and two linear layers. A basic block layer includes a 1-D convolutional layer and a max-pooling layer, which the activation between them is ReLU. The basic blocks are used for feature extraction, while the linear layers play a role in classification.

\begin{figure}[h]
	\centering
	\includegraphics[width=\linewidth]{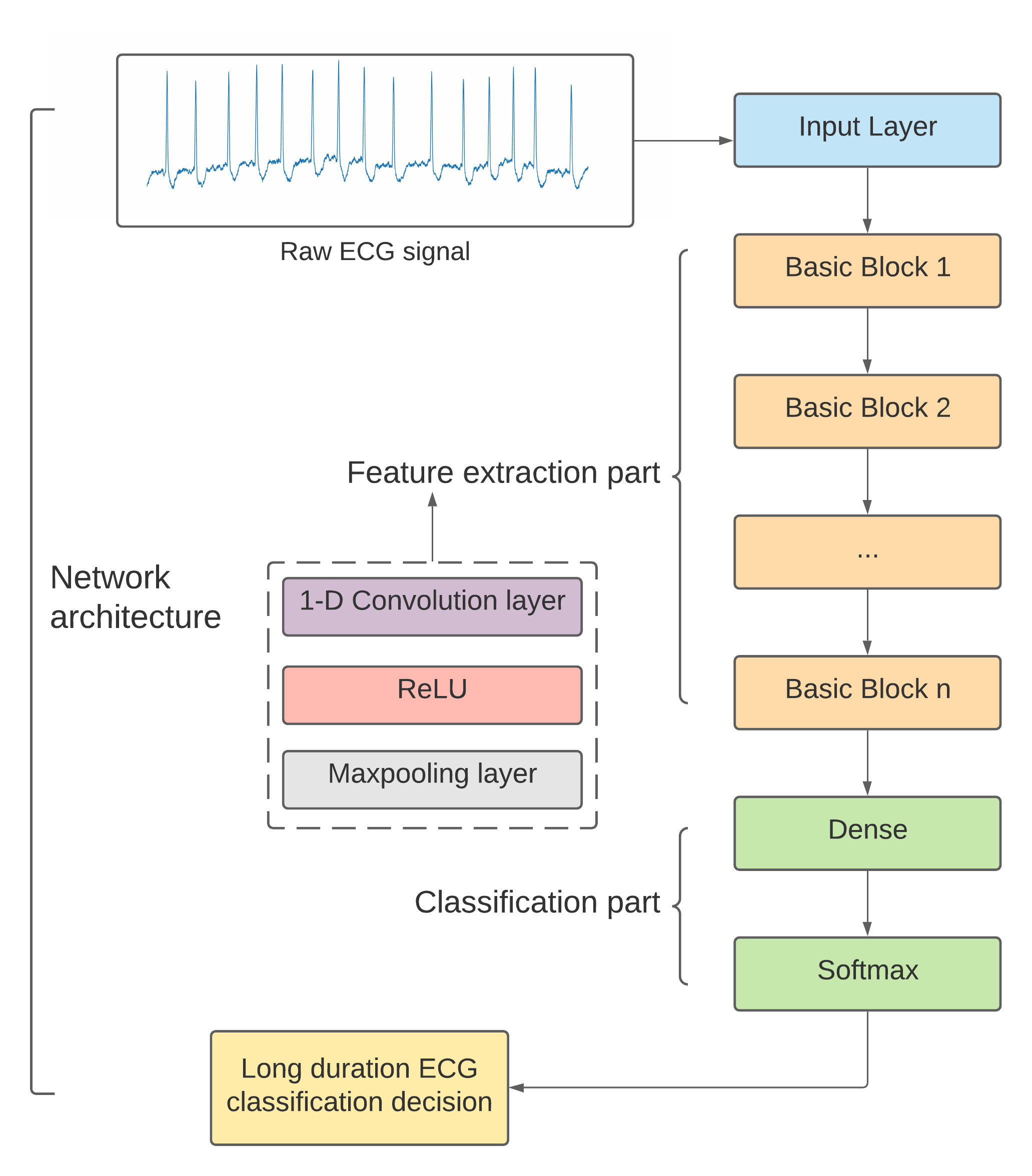}
	\caption{ECGNet structure}
	\label{fig:ECGNet}
\end{figure}
The input is raw and long-duration ECG signals, composed of 3,600 sampling points in the duration of 10 s. The network achieves end-to-end detection and inferences the classification output without manual feature extraction or feature segmentation and data processing of the original signals.

When we design the network structure, we make a tradeoff between the network size and accuracy. Finally, we decide that the number of basic blocks should be 7 in that such depth can produce considerable output. At the same time,  it retains the tiny size of the network parameters, which can be a consumption of memory and computation resources. Therefore, our proposed network is a 17 layer CNN, which will be discussed later in Sec. \uppercase\expandafter{\romannumeral4}.B.

\subsection{Adaptive Loss-aware MBNs Quantization }
Even though the depth of the proposed network architecture is minimal, the arrhythmia detection network still presents the problem of memory and power consumption due to the high bitwidth of weights. Because the importance of different layers varies, it is reasonable to quantize different layers with adaptive bitwidth, which can significantly reduce the average bitwidth of the network. We adopt the adaptive 1-D loss-aware multi-bit networks(MBNs) quantization method to help us realize the compression of the original proposed network.

Different from the quantization methods that minimize errors to reconstruct full precision weights, 1-D ALQ minimizes errors caused by quantization on the loss function. During the progress, neither gradient approximation nor full maintenance is involved \cite{qu2020adaptive}. After we train the full precision ECGNet, quantization process can be started. For the sake of improving the compression speed, parallel computing is introduced. For a vectorized weight $\boldsymbol{w}\in \mathbb{R}^{N\times 1}$, $\boldsymbol{w}$ is divided into  $m$ groups which are disjoint. Each group of weights is denoted by $\boldsymbol{w_k}\in \mathbb{R}^{n\times 1}$, where $N=n\times m$. Based on a binary basis, the quantized weights can be presented.

\begin{equation}
	\boldsymbol{w}_k=\sum_{i=1}^{I_k} \alpha_i \boldsymbol{\beta}_i = \boldsymbol{B}_k \boldsymbol{\alpha}_k,\quad \boldsymbol{\beta}_i\in \{-1,1\}^{n\times 1}
\end{equation}
We use $I_k$ to denote bitwidth of group $k$, and $\boldsymbol{B_k}$ represents the matrix forms of the binary bases. Therefore, we can define average bitwidth.
\begin{equation}
	\bar{I}=\frac{1}{m} \sum_{k=1}^m I_k
\end{equation}
Our target is to optimize the loss function of $\boldsymbol{w}_k$ so that it can help us reduce the average bitwidth $\bar{I}$, which is directly related to compression rate. Han et al. use a combination of pruning, quantization and  Huffman encoding to realize the multiple compression, which has excellent performance \cite{han2015deep}. Our 1-D ALQ combines pruning and quantization to achieve a better effect as well. Therefore, 1-D ALQ is composed of three steps, as shown in Fig. 3. Kernel weights are obtained according to the full-precision model parameters trained by the previously designed network. We perform flatten operation on the full-precision model parameters and then use the following three steps to achieve the final algorithm.
\begin{figure}[h]
	\centering
	\includegraphics[width=\linewidth]{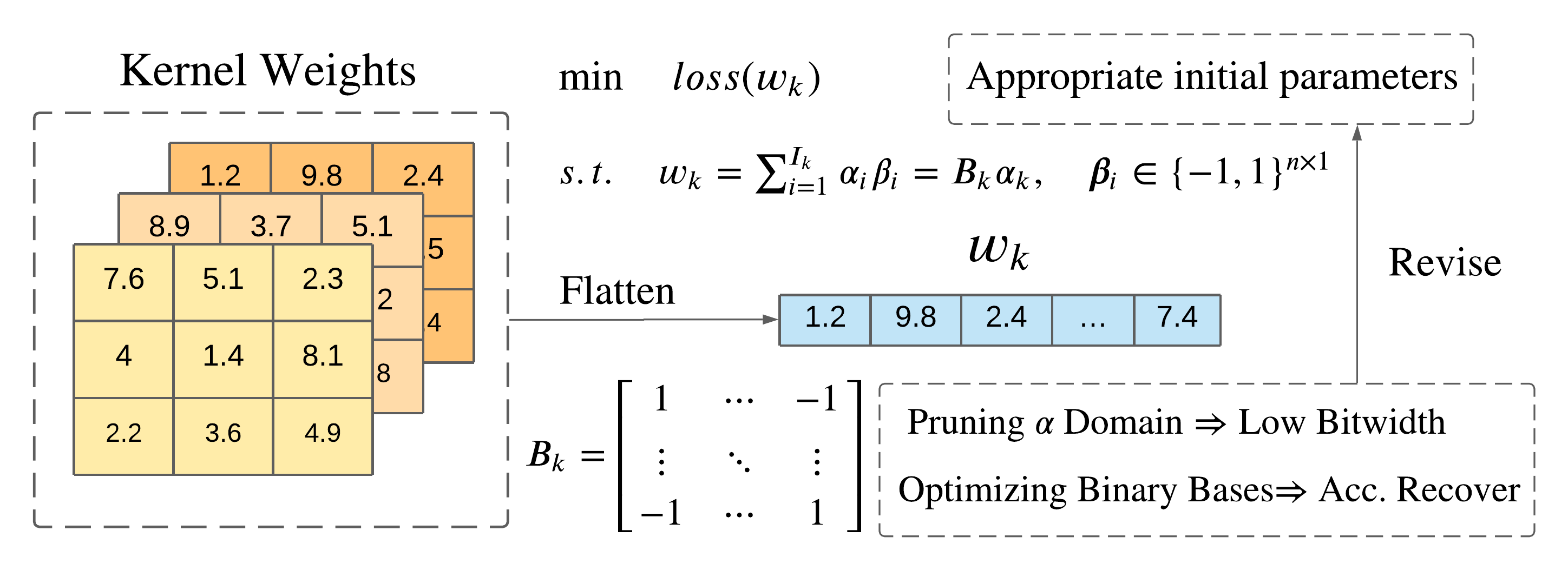}
	\caption{The algorithm of ALQ}
\end{figure}
\subsubsection{Initialize and Pruning}
In the beginning, we select a suitable initialization parameter to initialize the structure. Then we gradually reduce the average bitwidth $\bar{I}$ of the specific layer by pruning some coordinates which are of minimal importance in $\boldsymbol{\alpha}$ domain. Through pruning, many binary bases $\boldsymbol{\beta}_i$ are removed so that the consumption of the network decrease rapidly. However, regardless of detection performance, such reckless compression leads to poor 55.3\% accuracy in Sec. \uppercase\expandafter{\romannumeral4}.C. Although this step results in disappointing detection ability, we will recover such accuracy degradation introduced by the bitwidth reduction in the next step. 

\subsubsection{Optimize Binary Bases}
In this step, first, we fix the coordinates and search the binary base. After following the optimal value of binary bases, we fix the binary base to search the coordinates. This process takes the accuracy of detection as the standard and carries on several iterations. 
\subsubsection{Optimize Group Size $n$ and Other Initial Parameters}
After two steps above, the algorithm has been basically formed, but its compression rate and detection accuracy may not be that satisfactory. Therefore, we need to adjust some essential parameters according to the network structure. For example, we should select the appropriate maximum bitwidth $I_{max}$ and group size $n$. For the choice of maximum bitwidth, we should make a specific selection according to each layer's importance because the important layers tend to require more bitwidth to retain considerable accuracy. For group size $n$, 
Qu et al. consider that a group size from 32 to 512 achieves a good balance \cite{qu2020adaptive} while group size which is 16 can be better in our 1-D ALQ case.

\section{Experiment}
This section is the implementation of the designed network and compressing it using 1-D ALQ. It first introduces the dataset which used in the experiment. Then, it states the design details of the network structure and comparison with existing networks. In the Sec. \uppercase\expandafter{\romannumeral4}.C we realize the ALQ method experiment and compare our method with other quantization methods. Finally, we evaluate the ALQ method objectively and point out its drawbacks and possible future work.

\subsection{Dataset and Preprocessing}
ECG signals can be acquired from the MIT-BIH Arrhythmia database \cite{moody2001impact}, hosted at PhysioNet \cite{goldberger2000physiobank}. We extract 1,000 ECG signal fragments from this database sampled at a frequency of 360 Hz and lasted 10 seconds. 80\% of data is used as the training set while the rest is used as the test set. 

In the experiment, we normalize all the data, and then split the data index to ensure that the test set and training set are disjoint. In each experiment, all data were disordered and re-divided randomly to generate training sets and test sets for evaluating network performance.

\subsection{Convolutional Neural Network Topologies and  Diagnosis Performance}
Based on the design of the basic blocks and linear layer mentioned, we try different depths and numbers of convolution kernels. Considering the network's memory size and classification accuracy as the evaluation criteria, we find that more convolution kernels are needed in the shallow region. Therefore, ideally, the deeper and narrower the convolution kernel is perfectly suited to our needs. At the same time, we also need to consider the subsequent quantization process to adjust the network structure appropriately. Studies have shown that the accuracy of the detection hardly increases after the number of Basic Blocks exceeds seven.

Finally, after taking multiple factors into consideration, we constructed a network with seven basic blocks and two linear layers. The memory size of the network is 316.3KB, and the accuracy rate is 93.75\%. The detailed architecture design of the network is shown in Table 1.

\begin{table}[h]
	\caption{Network architecture}
		\begin{center}
			\setlength{\tabcolsep}{1.25mm}\begin{tabular}{llll}
				\hline
				Layer & Layer Name & Kernel $\times$ Unit & Other Layer Params  \\ 
				\hline
				1 & Conv1D & $16\times 8$ & ReLU, Strides=2, Padding=7 \\ 
				2 & MaxPooling1D & $8$ & Stride=4 \\ 
				3 & Conv1D & $12\times 12$ & ReLU, Strides=2, Padding=5 \\ 
				4 & MaxPooling1D & 4 & Stride=2 \\ 
				5 & Conv1D & $9\times 32$ & ReLU, Strides=1, Padding=4  \\
				6 & MaxPooling1D & 5 & Stride=2 \\
				7 & Conv1D & $7\times 64$& ReLU, Strides=1, Padding=3  \\
				8 & MaxPooling1D & 4 & Stride=2  \\ 
				9 & Conv1D & $5\times 64$ &  ReLU, Strides=1, Padding=2  \\
				10 & MaxPooling1D & $2$ & Stride=2 \\
				11 & Conv1D & $3\times 64$ &  ReLU, Strides=1, Padding=1 \\ 
				12 & MaxPooling1D & $2$ & Strides=2 \\
				13 & Conv1D & $3\times 72$ & ReLU, Strides=1, Padding=1  \\
				14 & MaxPooling1D & $2$ & Strides=2 \\ 
				15 & Flatten & - & -  \\
				16 & Dense & $1\times 216$ & ReLU, Dropout Rate=0.1 \\ 
				17 & Softmax & $1\times 17$ & -  \\
				\hline
			\end{tabular}
		\end{center}
\end{table}
As shown in Fig. 4, it is the normalized confusion matrix of the ECGNet above. As we can see, the accuracy of the detection is considerable that many of them reach 100\% accuracy, only a few are less than 90\% accuracy. For evaluating proposed network, the metrics are shown in (3)-(6), including the overall accuracy (\emph{OA}), specificity (\emph{Spe}) and sensitivity (\emph{Sen}).

\begin{equation}
	N = TP+TN+FP+FN
\end{equation}
\begin{equation}
	\emph{OA}=(\sum_{i=1}^{k}TP_i+TN_i)\cdot 100\%/N
\end{equation}
\begin{equation}
	\emph{Spe}=(\sum_{i=1}^{k}\frac{TN_i}{TN_i+FP_i})\cdot 100\%/k
\end{equation}
\begin{equation}
	\emph{Sen}=(\sum_{i=1}^{k}\frac{TP_i}{TP_i+FN_i})\cdot 100\%/k
\end{equation}
\begin{figure}[h]
	\centering
	\includegraphics[width=\linewidth]{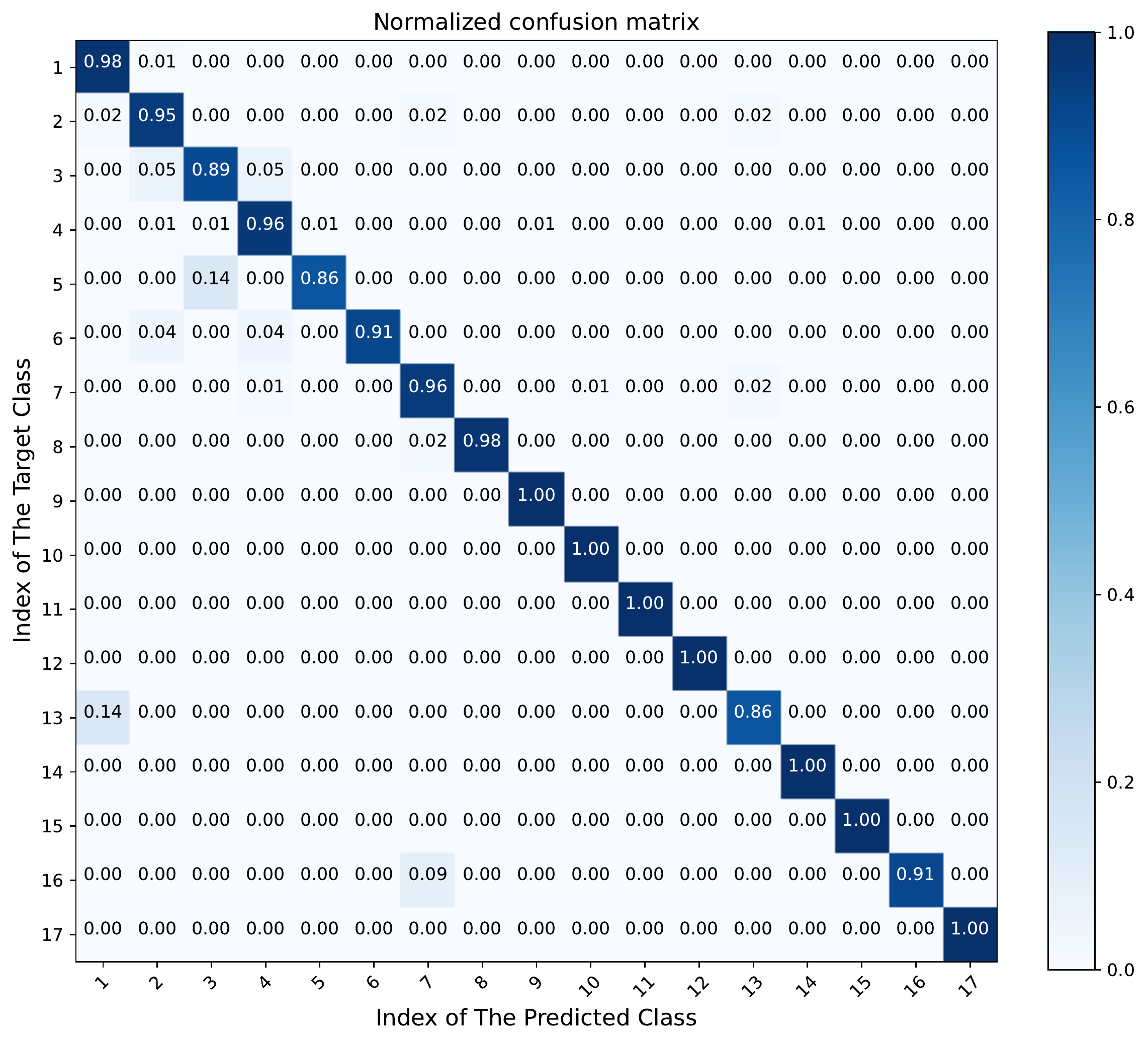}
	\caption{The normalized confusion matrix of ECGNet}
	\label{fig:Flowchart of the proposed framework}
\end{figure}

As Table 2 shown, comparing with three long-duration arrhythmia classifier \cite{rastegari2016xnor,yildirim2018arrhythmia,plawiak2018novel}, our network achieves an OA of 94.19\%, obviously improving the overall accuracy. Moreover, the overall accuracy will be improved again after our \cite{Serhani2020} quantization method.

\begin{table}[h]
	\caption{Performance of three existing long-duration arrhythmia classifiers and the proposed neural network }
	\label{tab:conf}
	\begin{minipage}{\columnwidth}
		\begin{center}
			\begin{tabular}{llll}
				\hline
				 & \emph{Sen}(\%) &\emph{Spe}(\%) & \emph{OA}(\%) \\
				\hline
				BWN Net \cite{rastegari2016xnor} & 90.85& 98.10 & 90.46\\
				DNN \cite{yildirim2018arrhythmia}&83.91&99.41&91.33\\
				P{\l}awiak\cite{plawiak2018novel} & 91.40 & 99.46 & 91.40 \\
				ECGNet (Proposed) &\textbf{94.19} & \textbf{99.67}&\textbf{93.5}  \\
				\hline
			\end{tabular}
		\end{center}
	\end{minipage}
\end{table}

\subsection{Compression Performance and Comparisons}

In this part, we realize the implementation compression of our ECGNet using adaptive loss-aware quantization as detailed in the last section. According to the study \cite{li2020arrhythmia}, the last layer and the second layer are more important layers for network structures like ours. Therefore, higher maximum bitwidth should be provided to these two layers. Other layers were offered with lower bitwidth. Table 3 illustrates the compression of the layers, respectively. As we expected, Softmax and Conv1D\_2 acquire higher average bitwidth, respectively 2.0000 and 1.9896.

\begin{table}[h]
	\caption{Compression results of layers}
	\label{tab:conf}
	\begin{minipage}{\columnwidth}
		\begin{center}
			\setlength{\tabcolsep}{1.25mm}\begin{tabular}{llll}
				\hline
				Layer & Average Bitwidth & Params & Memory \\ 
				\hline
				Conv1D\_1 & 1.2500 & 136 & 170 Bit \\ 
				Conv1D\_2 & 1.9896 & 1,164 & 2,316 Bit \\ 
				Conv1D\_3 & 1.7005 & 3,488 & 5,921 Bit \\ 
				Conv1D\_4 & 1.7095 & 14,400 & 24,617 Bit \\ 
				Conv1D\_5 & 1.4133 & 20,544 & 29,035 Bit \\ 
				Conv1D\_6 & 0.8545 & 12,352 & 10,555 Bit \\ 
				Conv1D\_7 & 0.8550 & 13,896 & 11,881 Bit \\ 
				Dense & 1.7422 & 13,888 & 24,196 Bit \\ 
				Softmax & 2.0000 & 1,105 & 2,210 Bit \\ 
				\hline
				\textbf{Total} & \textbf{1.3696} & 80,973 & \textbf{110,901 Bit = 13.538 KB} \\ 
				\hline
			\end{tabular}
		\end{center}
	\end{minipage}
\end{table}

As discussed above, adaptive bitwidths are given according to the characteristics and importance order of each layer. As seen from Fig. 5, some layers of the current ECG network architecture are more important and therefore a higher bitwidth is gained. This method optimizes the model error as much as possible and retains the inference precision as much as possible.

\begin{figure}[h]
	\centering
	\includegraphics[width=\linewidth]{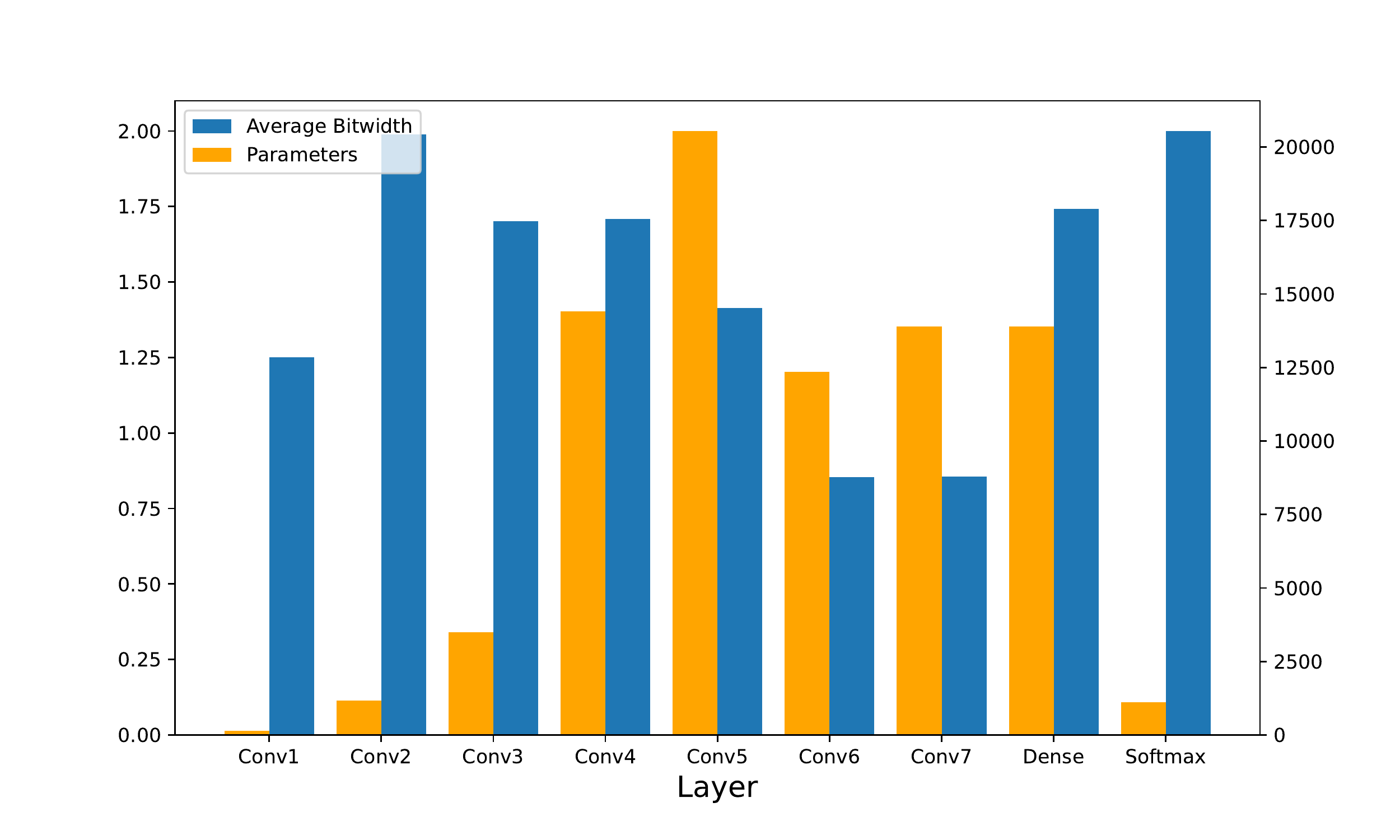}
	\caption{Distribution of the average bitwidth and the number of weights across layers}
\end{figure}

As Fig. 6 presented, with the increase of pruning rate, the decreasing amplitude of bitwidth is different, which indicates the method of adaptive bitwidth mentioned above. Moreover, we find the rate of decline is decreasing and it tends to stabilize at some value in the end.

\begin{figure}[h]
	\centering
	\includegraphics[width=\linewidth]{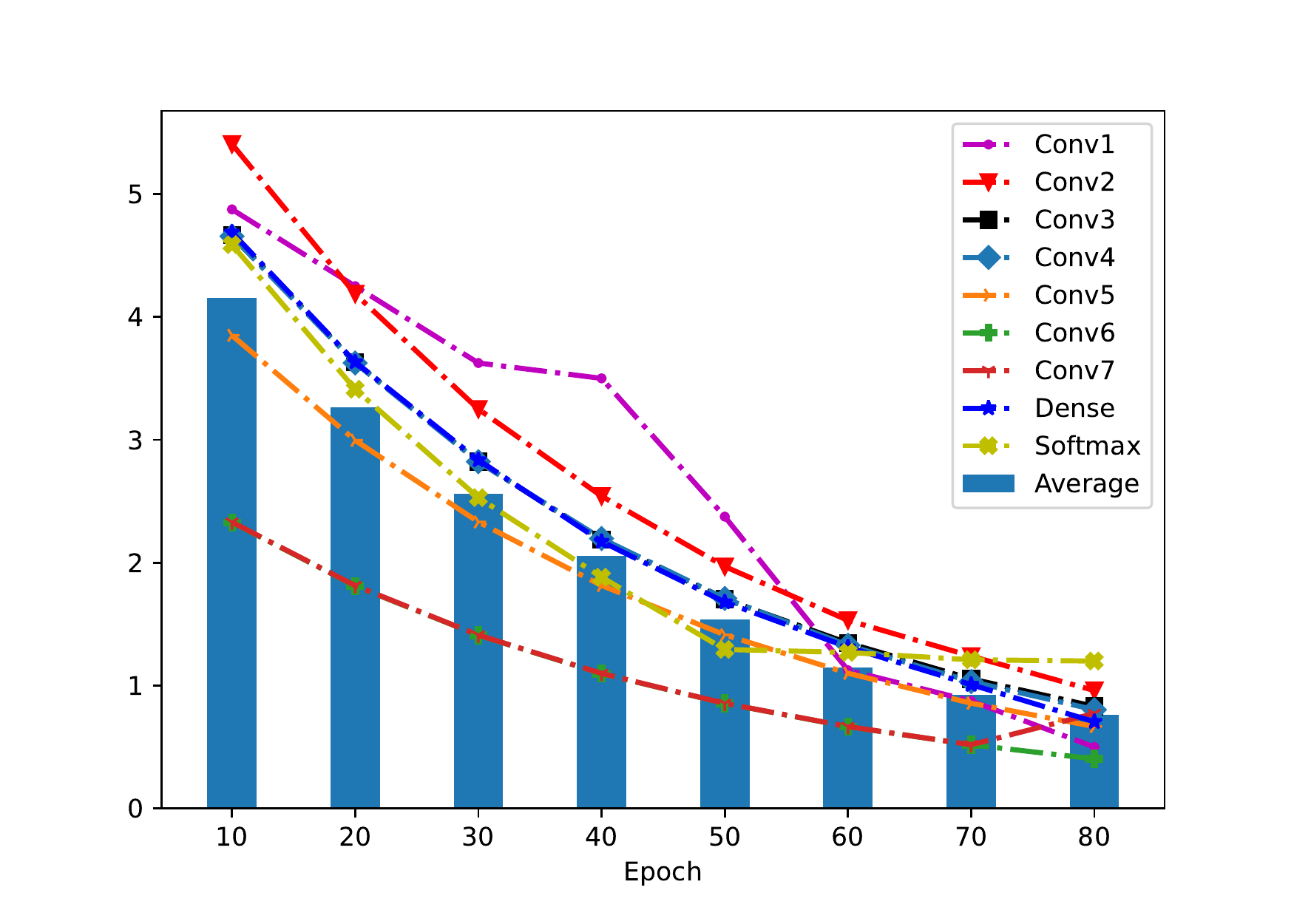}
	\caption{Change of bitwidth}
\end{figure}

Although we can greatly reduce the average bit width using this method, extreme reliance on this method can cause an unsatisfactory result, which we can observe in Fig. 7. As the bitwidth decreases, the loss function increases. To a certain extent, we can use the methods mentioned in the previous section to recover or even improve our accuracy. But when we go too far, the loss function increases dramatically, making our method ineffective and resulting in poor performance, which cannot be employed at all. Therefore, for our model, we choose appropriate parameters to improve the compression rate as much as possible while ensuring the accuracy rate of detection.

\begin{figure}[h]
	\centering
	\includegraphics[width=\linewidth]{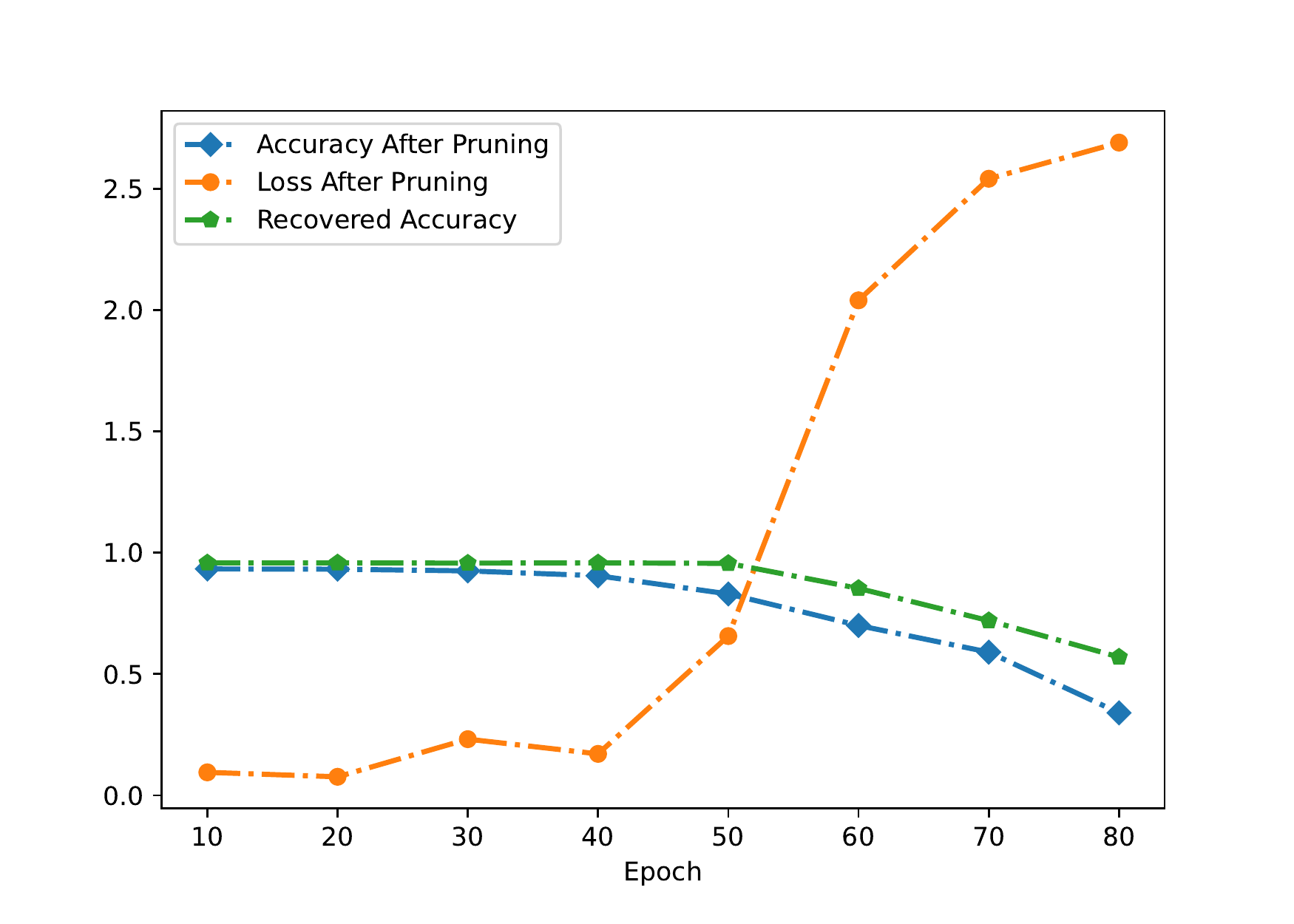}
	\caption{Accuracy and Loss}
\end{figure}

Table 4 compares the performance of several existing methods and the proposed quantization method. All of the quantization methods are employed on the aforementioned convolutional neural network to ensure that the comparison is fair. As shown in Table 4, our quantization method reaches the highest OA of 95.84\% with 2.34\% improvement compared to the unquantized network. Furthermore, the memory occupied by the compressed network parameters is just 13.54 KB, which is only 4.28\% of the original model's space occupation. Binary Connect \cite{park2018value}  compresses all weights to 1 bit while it has poor OA that is only 55.15\%, which is too imprecise to be able to be deployed as an arrhythmia classifier. P{\l}awiak et al. proposed DoReFaNet, reducing memory by 16 times and 10.67 times respectively for quantization to 2 and 3 bit. However, the degradation of accuracy is considerable as well. As for Accuracy-predominant \cite{li2020arrhythmia}, Memory-predominant \cite{li2020arrhythmia} and INQ \cite{huang2021efficient}, although these compression methods successfully achieve compression with low accuracy loss, their overall accuracy are 1.41\%,3.81\%, 3.07\% lower than proposed with prominent worse compression performance in compression rates, respectively. The comparison demonstrates that the proposed adaptive loss-aware quantization method has better performance, making it more suitable for deployment on hardware to achieve real-time heart rate detection.

\begin{table}[h]
	\caption{Performance of Six Methods and Our Proposed Adaptive Loss-Aware Quantization Method }
	\label{tab:conf}
	\begin{minipage}{\columnwidth}
		\begin{center}
			\begin{tabular}{lll}
				\hline
				Quantization method & \emph{OA}(\%) & Memory /Compression rate \\ 
				\hline
				Binary Connect \cite{park2018value} & 55.15\% & 9.84 KB/32.00 $\times$\\
				DoReFaNet (2 bit) \cite{plawiak2018novel}& 51.27\% & 19.67 KB/16.00$\times$ \\
				DoReFaNet (3 bit) \cite{plawiak2018novel}& 82.87\% & 29.50 KB/10.67$\times$\\	
				Accuracy-predominant \cite{li2020arrhythmia} & 94.43\% & 46.07 KB/6.83$\times$\\ 
				Memory-predominant \cite{li2020arrhythmia}& 92.13\% & 20.31 KB/15.50$\times$\\
				INQ \cite{huang2021efficient} & 92.76\% & 39.34 KB/8.11$\times$\\
				ALQ (Proposed)& \textbf{95.84\%}&	  \textbf{13.54 KB/23.36$\times$}\\	
				\hline
			\end{tabular}
		\end{center}
	\end{minipage}
\end{table}
\subsection{Discussion}
In summary, after the proposed quantization method, the network has better accuracy than other models and significantly reduce the memory occupation for hardware-friendly oriented. 

However, there are still some drawbacks introduced. After ALQ is quantized, our network weights have a coefficient, which is actually a design that adds extra overhead. Another problem is that the quantization of the activation layer is not considered, and relumax is not added, which may result in a loss of precision in hardware implementation. In future work, we have the chance to overcome these drawbacks, achieving a low-power arrhythmia detection ASIC chip that can be used on wearable devices.

\section{Conclusion}
Due to the limitation of memory and power, arrhythmia diagnosis on resource-constrained devices is remarkably difficult. In this paper, we propose an efficient convolutional neural network to detect long-duration ECG signal fragments with high recognition accuracy. Moreover, we adopt adaptive loss-aware quantization for multi-bit networks method to quantize our network, which realizes a $23.36\times$ compression rate and contributing improvement by 2.34\% of accuracy rate. The classification accuracy of our ECGNet in the MIT-BIH Arrhythmia database achieves 95.84\% while the required memory is compressed to 13.54 KB. In the future, we consider optimizing the quantization of activation layers and implementing this work on hardware platforms for real-time arrhythmia diagnosis.
\bibliographystyle{IEEEtrans}
\bibliography{ECG}

\end{document}